\documentclass{epl}
\usepackage{graphicx,amssymb}

\title{Non-analyticity in the distribution of conductances in quasi one
dimensional wires}
\shorttitle{Non-analyticity in the distribution ...}

\author{K. A. Muttalib\inst{1} \and P. W\"olfle\inst{2,3}
\and A. Garc\'{\i}a-Mart\'{\i}n\inst{2}
\and V. A. Gopar\inst{1} }
\shortauthor{K. A. Muttalib \etal}
\institute{
 \inst{1} Department of Physics - University of Florida,
P.O. Box 118440, Gainesville, FL 32611-8440, USA\\
\inst{2}Institut f\"ur Theorie der Kondensierten Materie -
Universit\"at Karlsruhe, P.O. Box 6980, D-76128 Karlsruhe, Germany\\
\inst{3}Institut f\"ur Nanotechnologie - Forschungszentrum Karlsruhe,
Germany.
}

\pacs{73.23.-b}{Electronic transport in mesoscopic systems}
\pacs{72.15.Rn}{Localization effects (Anderson or weak localization)}
\pacs{72.80.Ng}{Disordered solids}

\begin{document}

\maketitle

\begin{abstract}
We show that the distribution $P(g)$ of conductances $g$ of a
quasi one dimensional wire  has non-analytic behavior in the
insulating region, leading to a discontinuous derivative in the
distribution near
$g=1$. We give analytic expressions for the full distribution and
extract an approximate scaling behavior valid for different
strengths of disorder close to $g=1$.\\

\end{abstract}

It is well known that the probability distribution $P(g)$ of the
(dimensionless) conductances $g$ of a disordered conductor is
Gaussian in the deeply metallic regime and log-normal in the
deeply insulating regime \cite{review1}. It has been proposed
recently \cite{muttalib} that for a quasi one dimensional (1D)
wire with mean dimensionless conductance $\bar g \ll 1$, $P(g)$
has a sharp cut-off beyond $g=1$. To be sure, in one dimension,
when $g \leq 1$, $P(g)$ drops discontinuously to zero at $g=1$. In
higher dimensions the discontinuity will be smeared in some way,
but it is difficult to say how much of it will survive. Using the
saddle point method developed in \cite{muttalib}, we are able to
show here that nonanalytic behavior remains near $g=1$ at least in
quasi-1D systems. One should note that the exact results for the
mean and variance of the conductance, obtained within the
non-linear sigma model \cite{mirlin}, do not give any clue on the
abrupt change of $P(g)$ near $g=1$. The existence of the sharp
cutoff for all $\bar g \ll 1$ has recently been confirmed
numerically, while the exact shape of the distribution very close
to $g=1$ seems to be different from a log-normal distribution
\cite{garcia,markos,soukoulis,plerou,slevin}.

In the present work, we focus on the insulating region in quasi 1D
near $\vert g-1\vert \ll 1$,
in order to understand better the nature of the
unexpected sharp feature in the distribution.  For
simplicity, we will restrict our discussions to the unitary case
of broken time reversal symmetry. We propose from a simple generalization
of \cite{muttalib} that for large enough disorder the
distribution has a  non analyticity near $g=1$, giving rise to large
discontinuities  in its derivatives. We obtain e.g. that in the
insulating limit,
the leading contribution
to the distribution $P(g)$ at $\vert g-1\vert \ll 1$, is given by
\begin{equation}
P(g)\propto \cases{\Phi\left(\sqrt{\Gamma}(\nu-\frac{1}{2\Gamma})\right)
-\Phi\left(\sqrt{\Gamma}(\nu_1-\frac{1}{2\Gamma})\right)+C;
&$g\ge 1+\alpha$,\cr \Phi\left(\frac{1}{\sqrt{\Gamma}})\right)
-\Phi\left(\sqrt{\Gamma}(\nu_1-\frac{1}{2\Gamma})\right)+C;
&$g<1+\alpha$,}
\label{equ1}
\end{equation}
where $\Phi$ is the error function,
$\nu=\cosh^{-1}\sqrt{1/(g-1)}$, $\nu_1=\cosh^{-1}\sqrt{2/g}$, and
$\alpha=1/\cosh^2(3/2\Gamma)$ \cite{note1}. The disorder parameter
$\Gamma=\xi/L$, where $\xi$ is the localization length and $L$ is
the system size ($\Gamma \gg 1$ would correspond to metals and
$\Gamma \ll 1$ corresponds to insulators). The term $C$ is
independent of $g$, but depends on $\Gamma$. The discontinuity is
at $g=1+\alpha$. For the case considered here, $\Gamma \ll 1$ and
$\alpha \sim e^{-3/\Gamma}$. From~(\ref{equ1}) it follows that
$P^{\prime}=\upd P(g)/\upd g$ has a discontinuity at $g=1+\alpha$,
with $P^{\prime}\sim -e^{2/\Gamma}$ very large for $g \gtrsim
1+\alpha$ (growing exponentially with increasing disorder) and
$\vert P^{\prime}\vert \ll 1$ for $g < 1+\alpha$. Note that in the
limit $\Gamma \rightarrow 0$, the distribution would have an
essential singularity at $g=1$.

The above results are obtained assuming $\Gamma \ll 1$, describing the insulating
limit. As $\Gamma$ increases, but still in the insulating regime,
the dominant contribution remains the same with
renormalized parameters, and our results remain qualitatively valid.
In the metallic limit the singularity is absent,
giving rise to the possibility that the singularity disappears across the
crossover region ($\Gamma\sim 1/2$), where our current
approximations are not valid.

In addition to the sharp structure, our results suggest an
approximate scaling behavior for different strengths of disorder
near $g\sim 1$. As is evident from~(\ref{equ1}), in the expression for  the
ratio $P(g)/P(1)$, the dominant dependence on the disorder
parameter $\Gamma$ and on $g-1$ appears only in the
combination $\sqrt{\Gamma}(\nu-1/2\Gamma)$. Our numerical
results agree with this approximate scaling. We also find that the
distribution $P(g)$ for $g\lesssim 1$ is a slowly varying function
of $g$, which is approximately constant close to $g=1$.
This leads to an exponential distribution for $P(\ln g)$ in this
region, rather than the log-normal distribution proposed in
\cite{muttalib}, and agrees with various existing numerical
results \cite{markos,garcia}. For $\Gamma \ll1$, the numerical
results suggest that the exponential form crosses over to the
log-normal form for $g\ll 1$.

We now briefly discuss the details of the method which is based on a simple
generalization of \cite{muttalib}.
The probability distribution $p(\lambda)$ of the $N$ variables
$\lambda_i$, where $\lambda_i$ are related to the
transmission eigenvalues $T_i$ of an N-channel
quasi 1d wire by $\lambda_i=(1-T_i)/T_i$,  satisfy the well known DMPK
equation \cite{dmpk}, whose solutions in the metallic and insulating regimes can
be written in the general form \cite{beenakker}
\begin{equation}
p(\lambda)=\frac{1}{Z}\exp[-\beta H(\lambda)],
\label{equ2}
\end{equation}
where $Z=\int\prod_i \upd\lambda_i \exp[-\beta H]$ is a
normalizing factor independent of $\lambda_i$,
$H(\lambda)$ may be interpreted as the Hamiltonian
function of $N$ classical charges at positions $\lambda_i$,
and $\beta=2$ for the unitary case. Since the dimensionless
conductance is given by
\begin{equation}
g=\sum_{i}^{N}\frac{1}{1+\lambda_i},\label{equ3}
\end{equation}
the probability distribution $P(g)$ can be written as
\begin{equation}
P(g)=\frac{1}{Z}\int_{-\infty}^{\infty}
\frac{\upd\tau}{2\pi}\int_{0}^{\infty}\prod_{i=1}^{N}
\upd\lambda_i  \exp\left[i\tau (g-\sum_{i}^{N}\frac{1}
{1+\lambda_i})\right]p(\lambda).
\label{equ4}
\end{equation}
Following \cite{muttalib}, we define a ``Free energy''
\begin{equation}
F(\lambda)=\sum_{i}^{N}\frac{i\tau}{1+\lambda_i}+\beta H
\label{equ5}
\end{equation}
such that the distribution can be written as
\begin{equation}
P(g)=\frac{1}{Z}\int_{-\infty}^{\infty}
\frac{\upd\tau }{2\pi}e^{i\tau g}\int_{0}^{\infty}\prod_{i=1}^{N}
\upd\lambda_i  \exp\left[-F(\lambda)\right].
\label{equ6}
\end{equation}

It has been shown in \cite{muttalib} that separating out the lowest eigenvalue
$\lambda_1$ and
treating the rest of the eigenvalues as a continuum beginning at $\lambda_2$,
the
distribution can be obtained within a generalized saddle point approximation,
given by
\begin{equation}
P(g)=\int_{0}^{\infty}\upd\lambda_1
\int_{\lambda_1}^{\infty}\upd\lambda_2 e^{-S},
\label{equ7}
\end{equation}
where
\begin{equation}
S=-\frac{1}{2F^{\prime\prime}}
(g-F^{\prime})^2+F^0
\label{equ8}
\end{equation}
is the saddle point action, obtained from a saddle point free energy
\begin{equation}
F_{sp}=F^0+(i\tau)F^{\prime}+\frac{(i\tau)^2}{2}
F^{\prime\prime}.
\label{equ9}
\end{equation}
The insulating region in this model is given by $x_2 \gg 1$, and
also $x_2 \gg x_1$, where $\sinh^2 x_i=\lambda_i$. The saddle
point free energy terms in this limit were calculated to be
\cite{note2}
\begin{equation}
F^0(x_1,x_2)\approx 2\Gamma^2 x^2_2 - 6\Gamma x_2
+\Gamma x^2_1-\frac{1}{2} \ln (x_1 \sinh (2x_1));
\label{equ10}
\end{equation}
\begin{equation}
F^{\prime}\approx \frac{1}{\cosh^2 x_1};
\label{equ11}
\end{equation}
\begin{equation}
F^{\prime\prime}=-\frac{1}{\sinh^2 (2x_2)}
\left[\frac{1}{3}-\frac{1}{4x^2_2}+\frac{1}{\sinh^2 (2x_2)}\right].
\label{equ12}
\end{equation}
Since $1/2\vert F''\vert$ is exponentially large in the insulating regime,
 the saddle point solution
is given by putting
the coefficient of the term in the action $S$ equal to zero, namely $g=F'$,
which then gives
\begin{equation}
\cosh x_{1sp}=\frac{1}{\sqrt{g}}.
\label{equ13}
\end{equation}
As pointed out in \cite{muttalib}, the saddlepoint solution is
valid only for $g<1$ since
$\cosh x_1 \ge 1$, while the $g > 1$ region is determined by the boundary values
of $x_1=0$ and $x_2=2/\pi\Gamma$. While this is a good approximation on both sides
of $g=1$, the fact that the boundary of the saddle point solution is at
$g=1$ makes
it possible that the approximation is not accurate enough very close to
$g=1$. We
will show below that at the next level of approximation suggested in
\cite{muttalib}, the region close to $g=1$ can be better described by
saddle point
solutions valid on both sides. This improved approximation immediately leads
to the non-analyticity mentioned before.

It was shown in \cite{gopar} that while separating out the lowest eigenvalue gave
qualitatively good results both in the metallic and insulating regimes, it is
important to separate out one additional eigenvalue to obtain good agreement with
numerical and available exact results in the insulating and crossover regimes
because the separation between the eigenvalues becomes large. The
extension is straightforward, and the distribution $P(g)$ now has an additional
integral over $\lambda_3$, which is the beginning of the continuum that represents
the rest of the eigenvalues. It is clear that the insulating limit is characterized
by $x_3 \gg x_2 \gg 1$ for typical values of $x_i$, and that  $F'$ in this limit
will now be given by
\cite{gopar}
\begin{equation}
F^{\prime}\approx \frac{1}{\cosh^2 x_1}+\frac{1}{\cosh^2 x_2}.
\label{equ14}
\end{equation}
$F''$ now has the same expression as~(\ref{equ12}), with $x_2$ replaced by  $x_3$, while
$F^0$ is given by~(\ref{equ10}) with $x_1,x_2$ replaced by $x_2,x_3$ plus additional
terms involving $x_1$
\begin{equation}
F^0(x_1, x_2, x_3)=F^0(x_2,x_3)+\Gamma
x^2_1-\frac{1}{2}\ln(x_1\sinh (2x_1)).
\label{equ15}
\end{equation}

The saddle point solutions for $x_3$ and $x_2$ are now given by
$x_{3sp}=3/2\Gamma$, which is independent of $g$ and
\begin{equation}
\cosh x_{2sp}=\frac{1}{\sqrt{g-1/\cosh^2 x_1}},
\label{equ16}
\end{equation}
obtained again from $g=F'$.
Since $x_2 \gg 1$, the additional term in $F'$ is exponentially small, and
usually
negligible. However, close to $g=1$, this is the term that allows the
saddle point solution to be valid on both sides of $g=1$.
The fluctuation correction to the $x_2$ integral contributes an
additional factor $e^{2x_{2sp}}$, and
we are left with the final integral
\begin{equation}
P(g) \propto\int_{x_{1min}}^{x_{1max}}\upd x_1
e^{-\Gamma x^2_1+\ln x_1-\Gamma x^2_{2sp}+3x_{2sp}}.
\label{equ17}
\end{equation}
We have used finite limits in the integral
because the saddle point solution of $x_2$ must be real and also we require that
$x_{3sp} > x_{2sp}$.
This gives
$x_{1min}=Re[\cosh^{-1}\sqrt{1/(g-\alpha)}]$, $\alpha=1/\cosh^2(3/2\Gamma)$, and
$x_{1max}=\cosh^{-1}\sqrt{2/g}$. The crucial point is that the
restriction $x_{1min}$ is zero for all $g\ge 1+\alpha$, but
becomes finite for $g<1+\alpha$, with a discontinuous slope.
\emph{It is the universal nature of this constraint which will
eventually lead to the sharp changes in $P(g)$ at $g=1+\alpha$}.

\begin{figure}
\onefigure[scale=0.5]{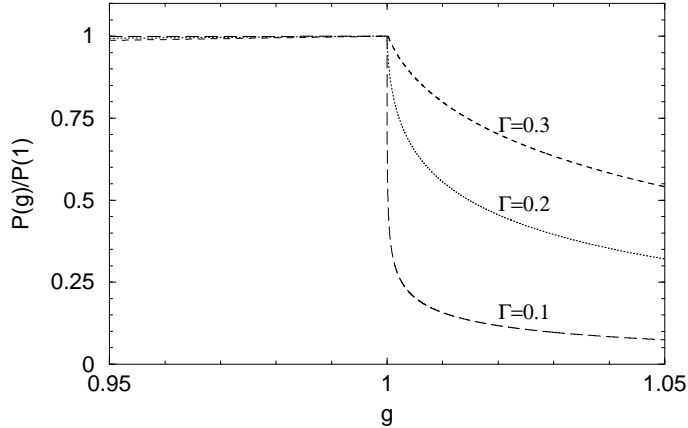}
\caption{$P(g)/P(1)$ from Eq.~(\ref{equ1}) for different values of $\Gamma$.}
\label{figure1}
\end{figure}

In the limit $g\rightarrow 1$ and $\Gamma \ll 1$, the upper limit of the integral
$x_{1max}\ll 1$. We can therefore neglect the $x^2_1$ term in~(\ref{equ17}).
Changing variables to $x_2$ and rewriting $\ln x_1$, we obtain a
simple integral
\begin{equation}
P(g)\propto \int_{\nu_1}^{x_{2max}} \upd x_2  e^{-\Gamma
x^2_2+ x_2}
\label{equ18}
\end{equation}
with $\nu_1$ as defined after~(\ref{equ1}), and $x_{2max}=\nu$ for $g>1+\alpha$, but
$3/2\Gamma$ for $g<1+\alpha$.
The integral immediately leads to Eq.~(\ref{equ1}). It is easy to see from~(\ref{equ1})
that the first derivative has a discontinuity
at $g=1+\alpha$ and is large, of order $e^{2/\Gamma}$ for $g > 1+\alpha$.
Figure~\ref{figure1} shows the sharp changes at
$g=1+\alpha$ according to Eq.~(\ref{equ1}).
Since $\nu_1$ in this region is only
weakly dependent on $g$, this also leads to an approximate scaling
behavior of $P(g)$ as a function of $z=\sqrt{\Gamma}(\nu-1/2\Gamma)$.

While the $g$-dependent contribution comes from the $x_2$ integral where
$x_{2sp} < x_{3sp}$, there is a contribution to $P(g)$ from integral over $x_2>x_{3sp}$. The
$x_3$ integral in this case is given by the boundary value at $x_3=x_2$, and the fluctuation
correction to the $x_2$ integral becomes of order unity. The resulting integral leads to a
$g$ independent but $\Gamma$ dependent term giving
$C\approx e^{3/\Gamma}[1-\Phi(2/\sqrt{\Gamma})]$.

In the above, we considered the case $\Gamma \ll 1$, where $x_3
\gg x_2 \gg 1$ and only the leading order terms were kept in the
free energy. For weaker disorder but still in the insulating side,
the next leading order terms will become important. First of all,
the $\ln x_2$ term arising from $F^0(x_2,x_3)$ in~(\ref{equ15})
will start to contribute. Second, the next leading order
contributions to $F^0$ will contribute some more $x_2$ and $\ln
x_2$ terms. In general, as long as the limit $x_3 \gg x_2 \gg 1$
remains valid, and the perturbative expansion of the free energy
remains a good approximation, the integral~(\ref{equ18}) can be
generalized to be of the form
\begin{equation}
P(g)\propto \int_{\nu_1}^{x_{2max}} \upd x_2 x^p_2 e^{-a \Gamma x^2_2+b x_2}
\label{equ19}
\end{equation}
where $a$, $b$ and $p$ are parameters which in
principle can be calculated within our framework by keeping higher order terms in the
free energy.
For $p=0$ we get back~(\ref{equ1}) with renormalized parameters. For finite integer $p$ the
integral can again be evaluated simply in terms of the error function.
For non-integer values of $p$, the integral is more complicated but it is clear that
the qualitative features of the results will remain the same. We expect that our
results will remain qualitatively valid near the crossover region on the insulating side
 ($\Gamma \lesssim 1/2$).

\begin{figure}
\onefigure[scale=0.5]{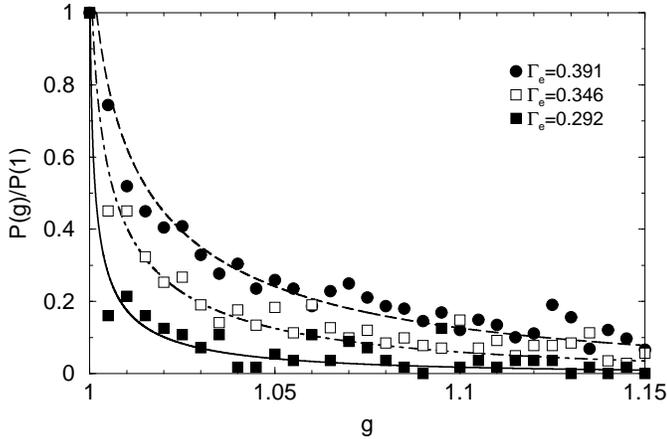}
\caption{Numerical data for $N=5$ with different disorder compared
with theoretical predictions using~(\ref{equ19}). Data for
$\bar{g}\approx 0.3$ were fit (solid line) with
$\Gamma_e=a\Gamma=0.292$ and $b=2$;  $\bar{g}\approx 0.4$ were fit
(dot-dashed line) with $\Gamma_e=0.346$ and $b=1.8$;
$\bar{g}\approx 0.5$ were fit (long dashed line) with
$\Gamma_e=0.391$ and $b=1.6$. In all plots, $p=2$.}
\label{figure2}
\end{figure}

\begin{figure}
\onefigure[scale=0.65,angle=-90]{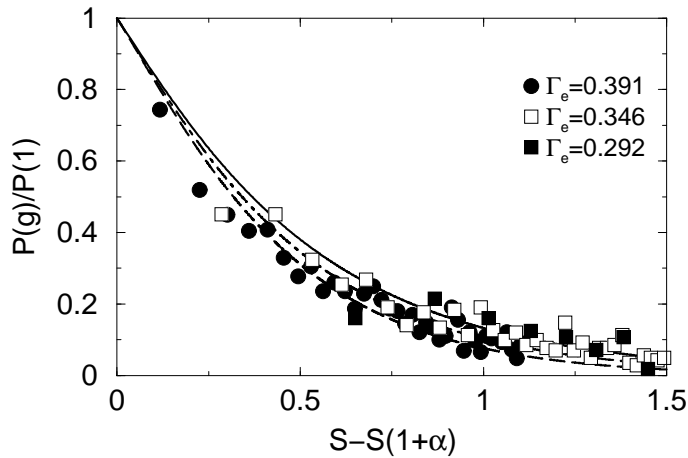}
\caption{Appropriately scaled data from Fig.~\ref{figure2} showing
approximate scaling as a function of
$s=\exp[-\sqrt{\Gamma_e}(\nu-b/2\Gamma_e]$, shifted by
$s(1+\alpha)$. The theoretical lines from Eq. (19) for different
$\Gamma_e$ (solid, dot-dashed and long-dashed lines for
$\Gamma_e=0.292$, $0.346$ and $0.391$, respectively) do not fall
exactly on each other, showing the approximate nature of the
scaling. } \label{figure3}
\end{figure}

In order to check  some of our approximate expressions, we have
calculated $P(g)$ numerically for a quasi 1D wire. Recent
techniques \cite{garcia} allow us to obtain reasonable statistics
for $N=5$. Unfortunately, it was not possible to obtain enough
data for either $\Gamma \ll 1$ or $1< g < 1.01$.
Nevertheless,~(\ref{equ19}) should be approximately valid in the
region numerically accessible to us and the data can be used to
check if the approximate scaling expected from ~(\ref{equ19})
holds qualitatively. In Fig.~\ref{figure2}, circles and squares
show the numerical results for different values of average
conductance $\bar g$. We use $p=2$, $\Gamma_e=a\Gamma$ and $b$ as
fitting parameters to fit the data for different $\bar g$
using~(\ref{equ19}) (for our range of parameters, the constant $C$
of~(\ref{equ1}) turns out to be negligible). Agreement with the
data is good for $g-1<0.15$.   In Fig.~\ref{figure3} the same data
are plotted as a function of the scaling variable
$s=\exp[-\sqrt{\Gamma_e}(\nu-b/2\Gamma_e]$ shifted by
$s(1+\alpha)$, taking into account the fact that the singularities
occur at $g=1+\alpha(\Gamma_e)$. Again the scaling is quite good.
We have checked that the scaling fails for larger values of
$\Gamma_e$, which is expected because it is close to the crossover
regime ($\Gamma\sim 1/2$) where our perturbative expansion of the
free energy starts to become invalid.

A non-analyticity in $P(g)$ was obtained in \cite{garcia} within
Random Matrix Theory (RMT), based on the assumption that
contributions from only the smallest two eigenvalues are important
in the insulating regime. All derivatives of $P(g)$ in such a
model diverge at $g=1$. However, a two-eigenvalue calculation
within DMPK (obtained by neglecting the continuum in our model)
gives rise to an essential singularity at $g=1$, so that there is
no discontinuity in the derivatives at that point. Therefore, it
is not simple to see if the trivial discontinuity in the
derivative mentioned in the introduction will persist beyond 1D.
We emphasize that our present results for quasi 1D are
fundamentally different from both of the above models: whereas all
derivatives of P(g) {\it diverge} for the RMT model and {\it
vanish} for the two-eigenvalue DMPK model, they are {\it finite}
for our quasi 1D system. Also, in contrast to the above cases, the
singularity for the quasi 1D system is not exactly at $g=1$, but
is shifted as a function of disorder. Finally, the RMT results do
not satisfy the approximate scaling of Fig.~\ref{figure3}.

We note that there is no phase transition in quasi 1D. It is therefore
important to ask if a non-analyticity in $P(g)$ in the absence of
a true phase transition violates any fundamental principle. Our framework can be
considered as an electrostatic problem in one dimension involving charges
with repulsive interactions and confinement potentials. Our free energy
and the density of charges are all analytic, ensuring that e.g. the total energy
will be an
analytic function of disorder. However, the conductance is a
complicated function of the charge distribution, and the non-analyticity appears
only in the distribution of the conductances where the charges have to satisfy
certain constraints. There is no restriction on the analyticity of such a
distribution.

In summary, we have shown that the distribution of conductances in
quasi 1D systems in the insulating regime has a non-analytic
behavior near $g=1$. The non-analyticity gives rise to very sharp
structures close to $g=1$, with finite discontinuities in its
derivatives. The presence of similar structures in higher
dimensions in numerically studied systems
\cite{markos,soukoulis,slevin} gives rise to the possibility that
such non-analyticity might be present in the conductance
distribution in higher dimensions as well, having important
consequences for the Anderson transition. Whether the
non-analyticity discussed here disappears abruptly at some
critical value of $\Gamma$, or smoothly, as $\Gamma$ is increased
beyond 1/2 remains to be clarified. It is conceivable that an
abrupt disappearance of the non-analyticity is a signature of the
Anderson transition.

KAM is grateful for support from the Alexander von Humboldt
foundation during his visit to U. Karlsruhe. Work of PW was
supported in part by SFB 195 of the Deutsche
Forschungsgemeinschaft (DFG). AGM was supported by the
Emmy-Noether program of the DFG under Grant No. Bu 1107/2-1. VAG
is grateful for support from CONACYT, M\'exico and hospitality at
the IPCMS, France.

%

%
%
%
%


\begin{thebibliography}{20}
\bibitem{note1}Eq. (1) is valid only for $|g-1|\ll 1$, but we
keep the more general definitions of $\nu$ and $\nu_1$ because
they will be needed later in ~(\ref{equ19}), expected to be valid
in an extended regime.
\bibitem{note2}Note that there is an error in the expression for $F^0$ in the
insulating limit in \cite{muttalib}; however this does not affect any of the results
obtained there.
\end{thebibliography}
\end{document}